\documentclass[article]{jss}
\usepackage[title]{appendix}
\usepackage[dvipsnames]{xcolor}
\usepackage{graphicx}
\usepackage{amsmath}
\usepackage{amssymb}
\usepackage{wrapfig}
\usepackage{mdframed}
\usepackage{amsbsy}
\usepackage{hyperref}
\usepackage{wrapfig}
\usepackage{animate}
\usepackage{algpseudocode}
\usepackage{algorithm}
\usepackage{algcompatible}
\usepackage{tcolorbox}
\usepackage[font={small,sf}]{caption}
\usepackage{algorithmicx}
\usepackage{algpseudocode}
\usepackage{float}
\usepackage{multicol}
\usepackage{caption}
\usepackage{amsfonts}
\usepackage{subfig}
\usepackage{color}
\usepackage{subfig}
\usepackage{url}
\usepackage[export]{adjustbox}[2011/08/13]
\usepackage{textpos}
\usepackage[percent]{overpic}
\DeclareMathOperator*{\argmin}{arg\,min}

\definecolor{mygreen}{rgb}{0,0.45,0}
\definecolor{mygray}{rgb}{0.99,0.99,0.99}
\definecolor{mymauve}{rgb}{0.58,0,0.82}

\usepackage{listings,lstautogobble}
\author{Charlie Wusuo Liu}
\title{\pkg{FLSSS}: A Novel Algorithmic Framework for Combinatorial Optimization Problems in the Subset Sum Family}

\Plainauthor{Charlie Wusuo Liu} %% comma-separated
\Plaintitle{A Novel Algorithmic Framework for Combinatorial Optimization Problems in the Subset Sum Family} %% without formatting
\Shorttitle{FLSSS} %% a short title (if necessary)

\Abstract{
  This article details the algorithmics in \pkg{FLSSS}, an \proglang{R} package for solving various subset sum problems. The fundamental algorithm engages the problem via combinatorial space compression adaptive to constraints, relaxations and variations that are often crucial for data analytics in practice. Such adaptation conversely enables the compression algorithm to drain every bit of information a sorted superset could bring for rapid convergence. Multidimensional extension follows a novel decomposition of the problem and is friendly to multithreading. Data structures supporting the algorithms have trivial space complexity. The framework offers exact algorithms for the multidimensional knapsack problem and the generalized assignment problem.
}
\Plainkeywords{subset sum, combinatorial optimization} %% without formatting
\Address{
  Charlie Wusuo Liu\\
  Boston, Massachusetts\\
  United States\\
  E-mail: \email{liuwusuo@bu.edu}
}
\graphicspath{{./}{../figure/}}

\begin{document}

\section{Introduction}\label{Introduction}
The Subset Sum problem is an NP-complete combinatorial optimization problem \citep{tssp}. Given a set of integers, it seeks a subset whose elements sum to a given target. Algorithms for solving the problem exist in a vast body of literature. These algorithms range over the exact and approximate approaches in both the deterministic and stochastic categories \citep{cbms,kkcx,acls,agafssp,afcr}. However, implementations of the algorithms are often inconvenient to access while the claimed performances remain in computational complexity analysis. The Subset Sum problem is formally defined in the integer domain, yet for data analytics, real numbers are typically the subjects. We often do not need or cannot even have a subset sum precisely equal the given target because of limited precision in floating-point arithmetics. 
%In practice, the constant factor hidden in the computational complexity notation could easily paralyze an algorithm either due to the algorithmic nature or bad engineering such as lack of hardware adaptation.

The package name of \pkg{FLSSS} stands for \textit{fixed-length subset sum solver}, the single function implemented in its first version. Algorithms in the package are meticulously implemented with many rounds of optimizations regarding both the mathematics and hardware adaptation for pushing speed limits. Solvers in the package differ from the mainstream definition of Subset Sum in the options of (i) restricting subset size, (ii) bounding subset elements, (iii) mining real-value sets with predefined subset sum errors, and (iv) finding one or more subsets in limited time. A novel algorithm for mining the one-dimensional Subset Sum (Section \ref{odfsss}) induced algorithms for the multi-Subset Sum (Section \ref{mss}) and the multidimensional Subset Sum (\mbox{Section \ref{mdss}}). The latter can be scheduled in a multithreaded environment, and the framework offers strong applications as exact algorithms to the multidimensional Knapsack (Section \ref{mkp}) and the Generalized Assignment problems (Section \ref{gap}) for solving to optimality. The package provides an additional functionality that maps reals to integers with controlled precision loss. These integers are further zipped non-uniformly in 64-bit buffers. Algebras (addition, subtraction, comparison) over compressed integers are defined through simple bit manipulations with virtually zero speed lags (Section \ref{icp}) relative to those over normal integers. Acceleration from the dimension reduction can be substantial.

Core algorithms in \pkg{FLSSS} are implemented in \proglang{C++}. Inexpensive preprocessing steps such as data assembling and reshaping are coded in \proglang{R}. The package employs \pkg{Rcpp} \citep{rcpp} APIs for getting memory addresses and size information of \proglang{R} objects. The basic multithreading template is taken from \pkg{RcppParallel} \citep{rcpppara}. Thread synchronization tools such as mutex and atomic classes are borrowed from Intel TBB library \citep{inteltbb} included in \pkg{RcppParallel}.

\section{One-dimensional Subset Sum}\label{odfsss}
The one-dimensional fixed-size Subset Sum algorithm (OFSSA) iterates two major steps: index hypercube contraction and subspacing. Assuming subset size $n$, the algorithm views a qualified subset as a point in an $n$-dimensional hypercube, each dimension of which consists of indexes of elements in a sorted superset. The contraction step compresses the hypercube to a smaller hyperrectangle. The subspacing step halves the hyperrectangle over a particular dimension and reshapes other dimensions accordingly. If certain dimension has only one index left, it is excluded and the problem reduces to an $(n-1)$-size Subset Sum.

Hypercube contraction and the data structure adaptive to it are the key algorithmics in OFSSA. The iteration of compression and subspacing falls in the branch-and-bound paradigm \citep{aamo}. Implementation of OFSSA focuses on maximizing the intensity of hypercube contraction and minimizing the waste of recalculations anywhere in the algorithm.

The one-dimensional variable-size Subset Sum can convert to a fixed-size problem via (i) doubling the superset size by padding zeros, and (ii) mining a subset of size equal to half of the new superset size. In practice, looping over all subset sizes is usually more efficient.

\subsection{Contraction}\label{Contraction}
A real-value superset of size $N$ is sorted and stored in an array
\begin{equation*}
\boldsymbol{x}=(x_0,x_1,\ldots,x_{N-1})=\big(\boldsymbol{x}(0),\boldsymbol{x}(1),\ldots,\boldsymbol{x}(N-1)\big)\;.
\end{equation*}
Given subset size $n$, we look for an integer array $\boldsymbol{i}=(i_0,i_1,\ldots,i_{n-1})$ such that 
\begin{equation}\label{ssrg}
\sum_{k=0}^{n-1}\boldsymbol{x}(i_k)\in[\text{MIN},\text{MAX}]\text{ , the subset sum range.}
\end{equation}
The index array $\boldsymbol{i}$ satisfies
\begin{equation}\label{ineq0}
\boldsymbol{x}(i_0) \leq \boldsymbol{x}(i_1) \leq\ldots \leq \boldsymbol{x}(i_{n-1})\;,
\end{equation}
\begin{equation}\label{ineq}
0\leq i_0 < i_1 <\ldots < i_{n-1}\leq n-1\;,
\end{equation}
\begin{equation}\label{ineq2}
i_k\in[k,\;N - n + k],\;k\in\{0,1,\ldots,n-1\}\;.
\end{equation}
%Inequalities \eqref{ineq0} and \eqref{ineq} reflect the lower (upper) bounds of $i_k$ sufficiently infer those of $\boldsymbol{x}(i_k)$. 
Equation \eqref{ineq2} outlines the $n$-dimensional hypercube where potential qualified subsets reside.

The contraction algorithm (i) finds the infima (greatest lower bounds) of $i_k$, $k$ $=$ 0 to $n-1$; (ii) finds the suprema (least upper bounds) of $i_k$, $k=n-1$ to 0; (iii) repeat (i) and (ii) until the infima and suprema become stationary. Note $k$ in (i) and (ii) proceeds in opposite directions. Let $l(i_k)$ and $u(i_k)$ be the current lower and upper bounds of $i_k$. 

\subsubsection{First index}
We inspect $l(i_0)$ first. The initial value of $l(i_0)$ equals 0 by  Equation \eqref{ineq2}. Our goal is to uplift $l(i_0)$ if possible. \mbox{Equation \eqref{ssrg}} implies $\boldsymbol{x}(i_0)\geq\text{MIN}-\sum_{t=1}^{n-1}\boldsymbol{x}(i_t)$. Notice the initial maxima of $\boldsymbol{x}(i_1),\ldots,\boldsymbol{x}(i_{n-1})$ are $\boldsymbol{x}(N-n+1),\ldots,\boldsymbol{x}(N-1)$, thus
\begin{equation}\label{i0}
\boldsymbol{x}(i_0)\geq\text{MIN}-\max\Big(\sum_{t=1}^{n-1}\boldsymbol{x}(i_t)\Big)=\text{MIN}-\sum_{t=N-n+1}^{N-1}\boldsymbol{x}(t)\;.
\end{equation}
Therefore, updating $l(i_0)$ is equivalent to solving the following optimization system :
\begin{equation}\label{i0up}
\begin{split}
l(i_0)\gets\min(\alpha)\hspace{1em}\text{subject to}\hspace{1em}\boldsymbol{x}(\alpha)\geq\text{MIN}-\sum_{t=N-n+1}^{N-1}\boldsymbol{x}(t)\;.
\end{split}
\end{equation}
System \eqref{i0up} updates the lower bound of $i_0$ to the index of the least element (in $\boldsymbol{x}$) that is no less than $\text{MIN}-\sum_{t=N-n+1}^{N-1}\boldsymbol{x}(t)$.

\subsubsection{Rest indexes}
For $k\in[1,\,n)$, the update of $l(i_{k-1})$ immediately triggers
\begin{equation*}
l(i_k)\gets\max\big(l(i_k),\,l(i_{k-1})+1\big)
\end{equation*}
because of Constraint \eqref{ineq}.

Similar to Inequality \eqref{i0}, we have
\begin{equation}\label{ik}
\boldsymbol{x}(i_k)\geq\text{MIN}-\max\Big(\sum_{t=0}^{k-1}\boldsymbol{x}(i_t)\Big)-\max\Big(\sum_{t=k+1}^{n-1}\boldsymbol{x}(i_t)\Big)\;.
%\geq\text{MIN}-\sum_{t=0}^{k-1}\boldsymbol{x}(i_t)-\sum_{t=k+1}^{n-1}\boldsymbol{x}\big(u(i_t)\big)\;.
\end{equation}
%In this inequality, the maximum of the sum 
%Apparently, $\max\big(\sum_{t=k+1}^{n-1}\boldsymbol{x}(i_t)\big)$
The sum $\sum_{t=k+1}^{n-1}\boldsymbol{x}(i_t)$ is maximized when $i_{k+1},\ldots,i_{n-1}$ equal their current individual maxima $u(i_{k+1}),\ldots,u(i_{n-1})$. To maximize the sum $\sum_{t=0}^{k-1}\boldsymbol{x}(i_t)$, we cannot simply assign $i_{0},\ldots,i_{k-1}$ to their current individual maxima, because Constraint \eqref{ineq} further upper-bounds $i_{0},\ldots,i_{k-1}$ with $i_k$. In fact, $i_{t}\leq i_k-(k-t)$ for any $t\in[0,\,k]$. Therefore,
\begin{equation*}
\max\Big(\sum_{t=0}^{k}\boldsymbol{x}(i_t)\Big)=\sum_{t=0}^{k}\boldsymbol{x}\Big(\min\big(u(i_t),\;i_k-k+t\big)\Big)\;,
\end{equation*}
and
\begin{equation}\label{likopt}
\begin{split}
&l(i_k)\gets\min(\alpha)\hspace{1em}\text{subject to}\\
&\hspace{2em}l(i_k)\leq\alpha\leq u(i_k)\;,\\
&\hspace{2em}\sum_{t=0}^{k}\boldsymbol{x}\Big(\min\big(u(i_t),\;\alpha-k+t\big)\Big)\geq\text{MIN}-\sum_{t=k+1}^{n-1}\boldsymbol{x}\big(u(i_t)\big)\;.\\
\end{split}
\end{equation}
Notice the left side of the second constraint is a non-decreasing function of $\alpha$, thus a brute-force solution to System \eqref{likopt} can be of (i) initializing $\alpha$ with the current $l(i_k)$, and (ii) incrementing $\alpha$ by 1 repeatedly until the second constraint turns true. If $\alpha=u(i_k)$ and the constraint is still unsatisfied, contraction fails and no qualified subsets would exist.

\subsubsection{Upper bounds}
Updating the lower and the upper bounds are symmetrical. The upper bound $u(i_k)$ is updated by
$u(i_k)\gets\min\big(u(i_k),\,u(i_{k+1})-1\big)$ first and then
\begin{equation}\label{uik}
\begin{split}
&u(i_k)\gets\max(\alpha)\hspace{1em}\text{subject to}\\
&\hspace{2em}l(i_k)\leq\alpha\leq u(i_k)\;,\\
&\hspace{2em}\sum_{t=k}^{n-1}\boldsymbol{x}\Big(\max\big(l(i_t),\;\alpha+t-k\big)\Big)\leq\text{MAX}-\sum_{t=0}^{k-1}\boldsymbol{x}\big(l(i_t)\big)\;.
\end{split}
\end{equation}
%We can assign the current $u(i_k)$ to $\alpha$ in the second constraint, decrement $i_k$ by 1 and stop once the inequality turns true.

\subsection{Contraction implementation}
The contraction algorithm is heavily optimized and far different from the narrative in \mbox{Section \ref{Contraction}}. These optimizations focus on Systems \eqref{likopt} and \eqref{uik}, and mainly consist of (i) decomposition of the $\min$, $\max$ operators, and (ii) an auxiliary quasi-triangle matrix of sums of consecutive elements in $\boldsymbol{x}$ for quick lookup.

\subsubsection{Decompose minimum function}
In the second constraint of System \eqref{likopt}, both $u(i_t)$ and $\alpha-k+t$ are increasing functions of $t$. When $t=k$, $\min\big(u(i_t),\,\alpha-k+t\big)=\alpha-k+t$ since $\alpha\leq u(i_k)$, thus we know $\alpha-k+t$ dictates the minimum function at the right end. Additionally, the discrete differential (slope) of $\alpha-k+t$ regarding $t$ is 1 and that of $u(i_t)$ is no less than 1 for every $t$, therefore, if $\alpha-k+t$ and $u(i_t)$ ever intersect, then $u(i_t)$ dictates the minimum function on the left of the intersection point and $\alpha-k+t$ dictates that on the right. The proof is trivial. Assuming $t^*$ the intersection point of $\alpha-k+t$ and $u(i_t)$, System \eqref{likopt} becomes
\begin{equation}\label{newsys}
\begin{split}
&l(i_k)\gets\min(\alpha)\hspace{1em}\text{subject to}\\
&\hspace{2em}l(i_k)\leq\alpha\leq u(i_k)\;,\;\text{(i)}\\
&\hspace{2em}\sum_{t=0}^{t^*-1}\boldsymbol{x}\big(u(i_t)\big)+\sum_{t=t^*}^{k}\boldsymbol{x}\big(\alpha-k+t\big)\geq\text{MIN}-\sum_{t=k+1}^{n-1}\boldsymbol{x}\big(u(i_t)\big)\;,\;\text{(ii)}\\
&\hspace{2em}\alpha-k+t^* \leq u(i_{t^*})\;.\;\text{(iii)}
%\;\; t=t^*,\ldots,k\;,\;\text{(iii)}\\
%&\hspace{2em}\alpha-k+t > u(i_t)\;,\;\;t=0,\ldots,t^*-1\;.\;\text{(iv)}
\end{split}
\end{equation}
If $t^*=0$, then $\alpha-k+t$ and $u(i_t)$ have no intersection and the first sum term in \eqref{newsys}(ii) is ignored. Our next goal is to find an appropriate $t^*$. We will see knowing $t^*$ brings considerable computational advantage.

\begin{figure}[t!]
	\begin{overpic}[grid=0,width=\textwidth,tics=2]{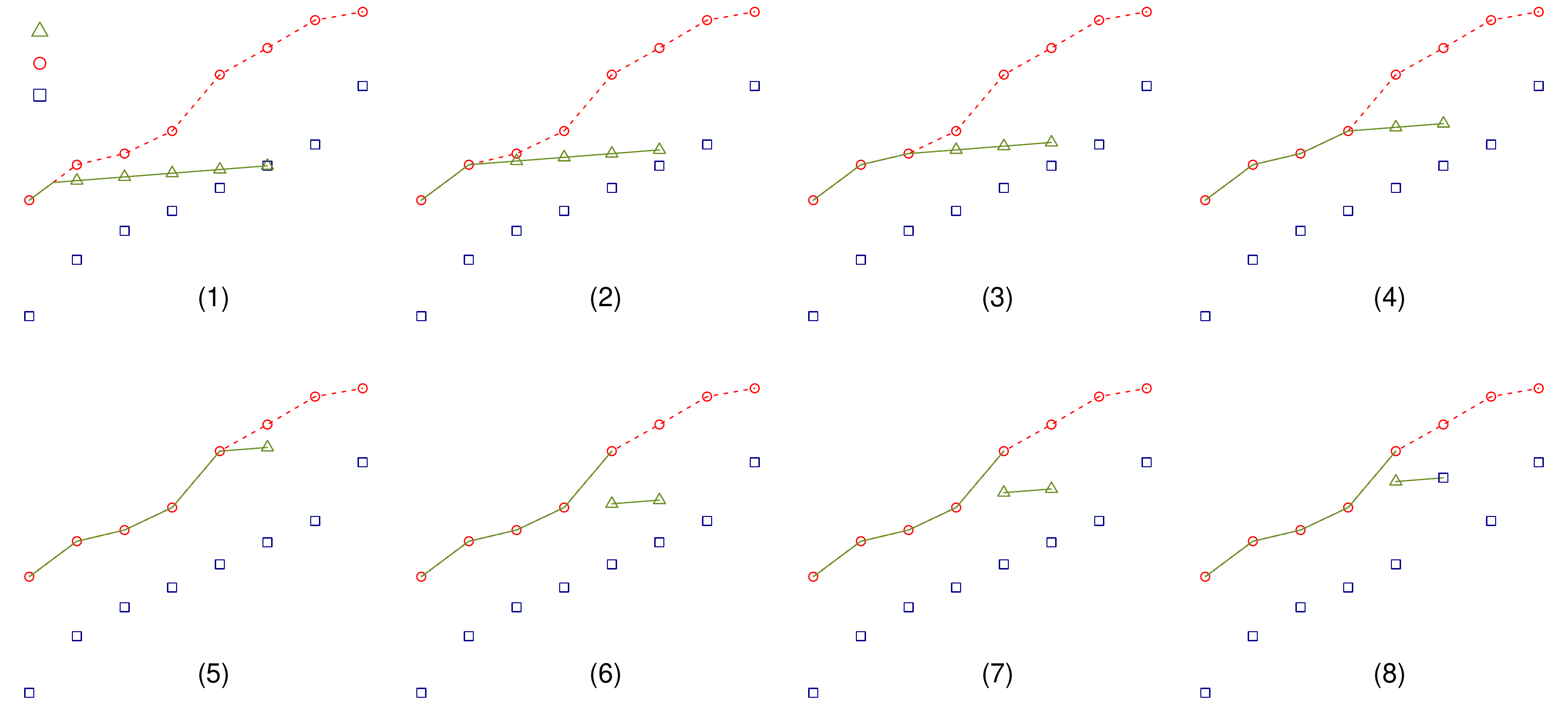}
		\put(4,40.5){\small$u(i_t)$}
		\put(4,38.5){\small$l(i_t)$}
		\put(4,42.5){\small$\alpha-k+t$}
		\put(18,33){\scriptsize$l(i_k)$}
		\put(14,31){\scriptsize$l(i_{k-1})$}
		\put(12,35){\scriptsize$l(i_{k})-1$}
		\put(8,17){$u(i_{t^*})$}
	\end{overpic}
	\caption{Panels (1)-(5) update $t^*$ by Inequality \eqref{eq11} using linear search. Panels (6)-(8) update $l(i_k)$ by Inequality \eqref{eq12} using linear search.}\label{fg}
\end{figure}
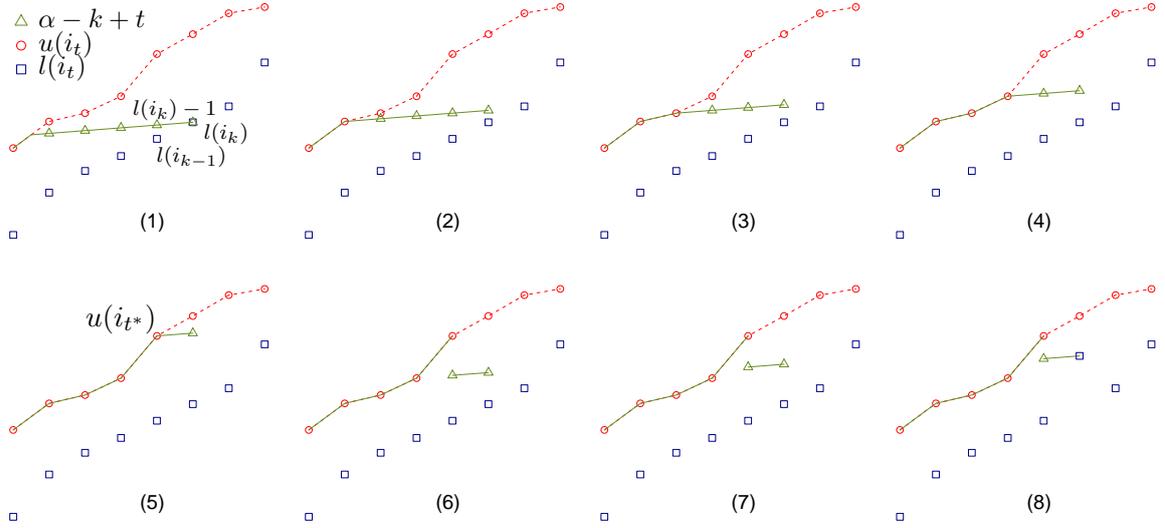

For convenience, let
\begin{equation}\label{f}
f(t^*,\,\alpha)=\sum_{t=0}^{t^*-1}\boldsymbol{x}\big(u(i_t)\big)+\sum_{t=t^*}^{k}\boldsymbol{x}\big(\alpha-k+t\big)\;.
\end{equation}
Notice $f$ is a non-decreasing function of $\alpha$. Constraint \eqref{newsys}(iii) implies $\alpha\leq u(i_{t^*})+k-t^*$, thus we have the following constraint:
\begin{equation}\label{eq11}
\max(f)=\sum_{t=0}^{t^*-1}\boldsymbol{x}\big(u(i_t)\big)+\sum_{t=t^*}^{k}\boldsymbol{x}\big(u(i_{t^*})+t-t^*\big)\geq\text{MIN}-\sum_{t=k+1}^{n-1}\boldsymbol{x}\big(u(i_t)\big)\;.
\end{equation}
Function $\max(f)$ is a non-decreasing function of $t^*$, which can be easily proved by valuing the discrete differential regarding $t^*$. We update $t^*$ using linear search. The initial value of $t^*$ was found in the prior iteration of $k-1$, and is incremented by 1 repeatedly until \mbox{Inequality \eqref{eq11}} becomes true. Here is a computing shortcut: except for $\sum_{t=t^*}^{k}\boldsymbol{x}\big(u(i_{t^*})+t-t^*\big)$ (see Section \ref{qtm} for its update process), sum terms in \mbox{Inequality \eqref{eq11}} are not updated by summations but by adding (subtracting) the incoming (outgoing) elements when $t^*$ is incremented.  The updated $t^*$ is the least $t^*$ to let \mbox{Inequality \eqref{newsys}(ii)} possibly be satisfied for some $\alpha$. \mbox{Inequality \eqref{newsys}(ii)} is then rewritten as
\begin{equation}\label{eq12}
\sum_{t=t^*}^{k}\boldsymbol{x}\big(\alpha-k+t\big)\geq\text{MIN}-\sum_{t=k+1}^{n-1}\boldsymbol{x}\big(u(i_t)\big)-\sum_{t=0}^{t^*-1}\boldsymbol{x}\big(u(i_t)\big)\;.
\end{equation}
We initialize $\alpha\gets\max\big(l(i_k),\,u(i_{t^*-1}) + (k-t^*+1)\big)$, and repeatedly increment $\alpha$ by 1 until Inequality \eqref{eq12} turns true. The resulting $\alpha$ becomes the new $l(i_k)$. Figure \ref{fg} demonstrates a visual explanation for updating $t^*$ and $l(i_k)$. Decomposing the $\max$ operator in System \eqref{uik} follows the same rationale.

\subsubsection{Quasi-triangle matrix}\label{qtm}
\begin{figure}[t!]
	\includegraphics[width=\linewidth]{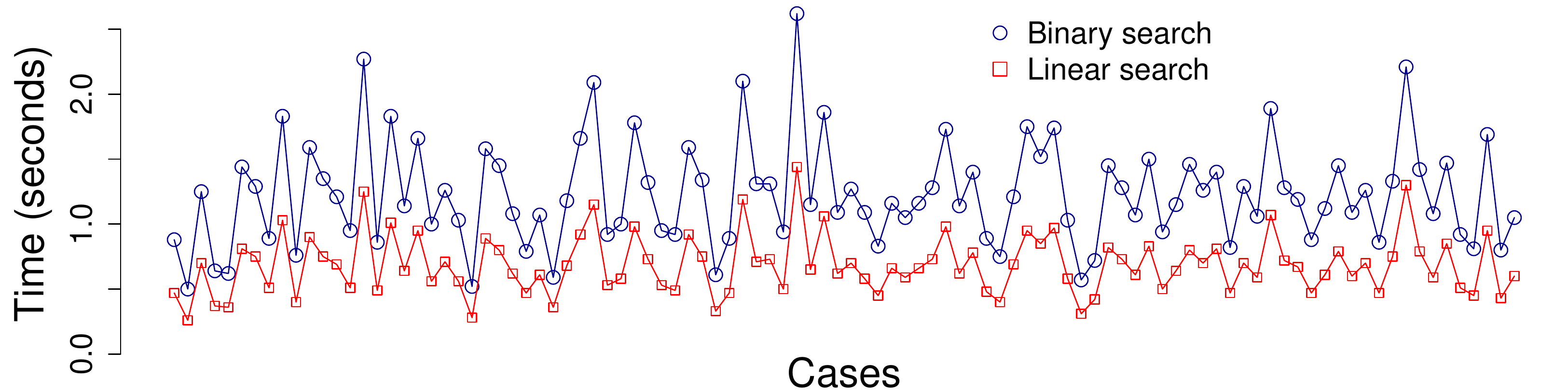}
	\caption{Supersets sizes = 1000, subsets sizes = 100, subset sum error thresholds =  0.0001, requested number of subsets $\geq$ 10, g++ '-O2' compile, Intel(R) i7-4770 CPU @ 3.40GHz, Windows 7. Preprocessing time included. Each of the 100 supersets contains uniforms in [0, 1000000]; the corresponding target sum is the sum of elements of a subset sampled at random. Linear search yields about 1.8x acceleration.}\label{lbt}
\end{figure}

The first sum in Inequality \eqref{eq12} and the second sum in \eqref{eq11} both add consecutive elements in $\boldsymbol{x}$. We construct the following matrix
\begin{equation}
\mathcal{M}=\left(\begin{array}{ccccc}
\boldsymbol{x}(0) & \sum_{t=0}^1\boldsymbol{x}(t) & \sum_{t=0}^2\boldsymbol{x}(t) & \ldots & \sum_{t=0}^{n-1}\boldsymbol{x}(t)\\
\boldsymbol{x}(1) & \sum_{t=1}^2\boldsymbol{x}(t) &\sum_{t=1}^3\boldsymbol{x}(t) & \ldots & \vdots\\
\vdots & \vdots & \vdots & &\sum_{t=N-n+1}^N\boldsymbol{x}(t)\\
\boldsymbol{x}(N-3) & \sum_{t=N-3}^{N-2}\boldsymbol{x}(t) & \sum_{t=N-3}^{N-1}\boldsymbol{x}(t)\\
\boldsymbol{x}(N-2) & \sum_{t=N-2}^{N-1}\boldsymbol{x}(t)\\
\boldsymbol{x}(N-1)
\end{array}\right)
\end{equation}
for fast looking up the sums. For instance, $\sum_{t=t^*}^{k}\boldsymbol{x}\big(\alpha-k+t\big)=\mathcal{M}[\alpha-k+t^*,\,k-t^*]$. This matrix is constructed once and used for every contraction until the qualified subsets are found. Because each column of $\mathcal{M}$ is in ascending order, updating $l(i_k)$ by \mbox{Inequality \eqref{eq12}} can also use binary search. However, simulations show binary searches here usually have lower performance due to CPU caching mechanisms \citep{tlp}. 
%\mbox{Code \ref{lrBiSearchTime}}
The following code
compares the time costs of mining a hundred supersets using binary search and linear search for contraction. \mbox{Figure \ref{lbt}} shows the results.

%\renewcommand{\lstlistingname}{Code}
%\begin{lstlisting}[caption={Compare computing speeds of using linear and binary search in contraction.}, label={lrBiSearchTime}]
%set.seed(42)
%lrtime = numeric(100)
%bitime = numeric(100)
%for(i in 1L : 100L)
%{
%  superset = sort(runif(1000, 0, 1e6)) # Generate a superset.
%  len = 100
%  target = sum(sample(superset, len)) # Make target sum.
%  me = 1e-4
%  lrtime[i] = system.time({FLSSS::FLSSS(len = len, v = superset, target = target, ME = me, solutionNeed = 10, useBiSrchInFB = F)})['elapsed']
%  bitime[i] = system.time({FLSSS::FLSSS(len = len, v = superset, target = target, ME = me, solutionNeed = 10, useBiSrchInFB = T)})['elapsed']
%}
%\end{lstlisting}

\begin{CodeChunk}
\begin{CodeInput}
R> set.seed(42)
R> lrtime = numeric(100)
R> bitime = numeric(100)
R> len = 100
R> me = 1e-4
R> for(i in 1L : 100L)
+  {
+    superset = sort(runif(1000, 0, 1e6))
+    target = sum(sample(superset, len))
+    lrtime[i] = system.time({FLSSS::FLSSS(len = len, v = superset, 
+      target = target, ME = me, solutionNeed = 10, 
+      useBiSrchInFB = FALSE)})['elapsed']
+    bitime[i] = system.time({FLSSS::FLSSS(len = len, v = superset,
+      target = target, ME = me, solutionNeed = 10, 
+      useBiSrchInFB = TRUE)})['elapsed']
+  }
R> mean(bitime / lrtime)
\end{CodeInput}
\end{CodeChunk}
\begin{CodeChunk}
\begin{CodeOutput}
[1] 1.790667
\end{CodeOutput}
\end{CodeChunk}

It is easy to see the infima $l(i_t)$ and suprema $u(i_t)$ will become stationary after finite contractions because $l(i_t)\leq u(i_t)$. The uniqueness of the stationed infima and suprema remains to be proved. \pkg{FLSSS} provides a function \code{z\_findBound()} for examining the concept. 
%In Code \ref{contractExample}, 
In the following example code,
the first call to \code{z\_findBound()} contracts a 10-dimensional hypercube starting with the infima. The second call to \code{z\_findBound()} starts with the suprema. Both calls converge to the same hyperrectangle.

%\renewcommand{\lstlistingname}{Code}
%\begin{lstlisting}[caption={Contract a 10-dimensional hypercube.}, label={contractExample}]
%# A superset of 10 elements.
%x = c(14, 60, 134, 135, 141, 192, 199, 203, 207, 234)
%MIN = 813
%MAX = 821
%lit = as.integer(c(1, 2, 3, 4, 5))  # R indexes start from 1.
%uit = as.integer(c(6, 7, 8, 9, 10)) # Hypercube upper bounds.
%
%hyperRectangle = FLSSS:::z_findBound(len = 5, V = as.matrix(x), target = (MIN + MAX) / 2, me = (MAX - MIN) / 2, initialLB = lit, initialUB = uit, UBfirst = FALSE)
%print(hyperRectangle[-1])
%# [[1]]
%# [1] 1 3 5 6 8
%# [[2]]
%# [1] 3 6 8 9 10
%
%hyperRectangle = FLSSS:::z_findBound(len = 5, V = as.matrix(x), target = (MIN + MAX) / 2, me = (MAX - MIN) / 2, initialLB = lit, initialUB = uit, UBfirst = TRUE)
%print(hyperRectangle[-1])
%# [[1]]
%# [1] 1 3 5 6 8
%# [[2]]
%# [1] 3 6 8 9 10
%\end{lstlisting}

\begin{CodeChunk}
\begin{CodeInput}
R> x = c(14, 60, 134, 135, 141, 192, 199, 203, 207, 234)
R> MIN = 813
R> MAX = 821
R> lit = as.integer(c(1, 2, 3, 4, 5))
R> uit = as.integer(c(6, 7, 8, 9, 10))
R> hyperRectangle1 = FLSSS:::z_findBound(len = 5, V = as.matrix(x), 
+    target = (MIN + MAX) / 2, me = (MAX - MIN) / 2, initialLB = lit, 
+    initialUB = uit, UBfirst = FALSE)
R> hyperRectangle2 = FLSSS:::z_findBound(len = 5, V = as.matrix(x), 
+    target = (MIN + MAX) / 2, me = (MAX - MIN) / 2, initialLB = lit, 
+    initialUB = uit, UBfirst = TRUE)
R> hyperRectangle1[-1]; hyperRectangle2[-1]
\end{CodeInput}
\end{CodeChunk}
\begin{CodeChunk}
\begin{CodeOutput}
[[1]]
[1] 1 3 5 6 8
[[2]]
[1] 3 6 8 9 10
[[1]]
[1] 1 3 5 6 8
[[2]]
[1] 3 6 8 9 10
\end{CodeOutput}
\end{CodeChunk}

\subsection{Subspacing}

Some previous versions of \pkg{FLSSS} (i) select a dimension of the hyperrectangle resulted from contraction, (ii) fix the value of that dimension, (iii) reduce the dimensionality of the problem by 1 and (iv) execute contraction again. In step (i), it chooses the dimension having the least domain width so to produce the fewest branches. These steps are a mixture of depth-first and best-first searches. 

\begin{figure}[t!]
	\includegraphics[width=\linewidth]{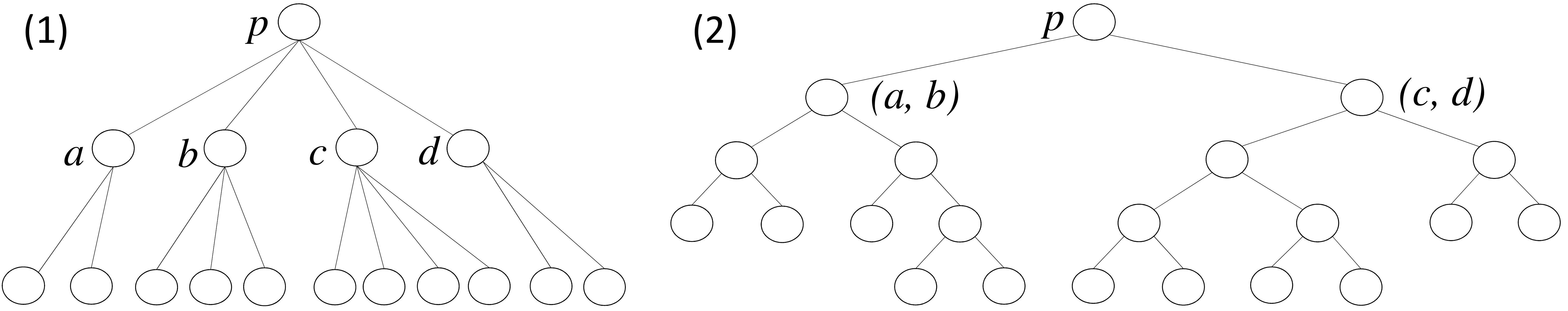}
	\caption{Panel (1) shows the variable-subspacing (VS) method. Assuming the narrowest dimension of hyperrectangle $p$ has width 4, then subspacing $p$ would produce 4 child hyperrectangles $a, b, c, d$. On the other hand, binary-subspacing (BS) halves $p$. If contracting \textit{a}, \textit{b} would both fail, VS has to contract both $a$ and $b$ to know, yet BS only has to contract \textit{(}$a,b$\textit{)}. For BS, \textit{c} or \textit{d} are decedents of \textit{(}$c,d$\textit{)}, a smaller hyperrectangle than $p$ that gives birth to \textit{c} and \textit{d} in VS.}\label{fg2}
\end{figure}

\begin{algorithm}
	\caption{Subspacing}\label{sbs}
	\begin{overpic}[grid=0,width=\textwidth,tics=2]{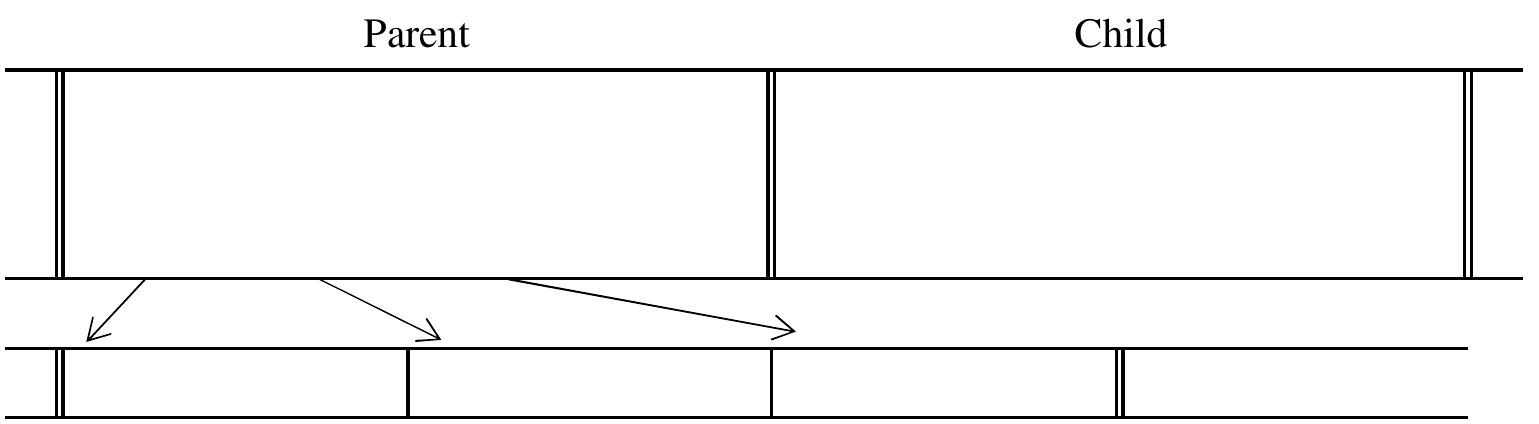}
		\put(8,20.5){$\beta$}
		\put(6,15.5){MIN}
		\put(8,10.5){$l$}
		\put(18,20.5){$\kappa$}
		\put(16,15.5){MAX}
		\put(18,10.5){$u$}
		\put(28,20.5){$n$}
		\put(24,15.5){$\sum\limits_{t=0}^{n-1}\boldsymbol{x}\big(l(i_t)\big)$}
		\put(28,10.5){$u^{\prime}$}
		\put(38,20.5){$n_z$}
		\put(37,15.5){$\sum\limits_{t=0}^{n-1}\boldsymbol{x}\big(u(i_t)\big)$}
		\put(38,10.5){$S(u)$}
		\put(6,1.5){$l(i_0),\ldots,l(i_{n-1})$}
		\put(28,1.5){$u(i_0),\ldots,u(i_{n-1})$}
		\put(52,1.5){$u(i_0),\ldots,u(i_{\kappa})$}
	\end{overpic}
	A hyperrectangle object in stack consists of twelve parameters explained in the following algorithm.
	
	\textbf{LEFT BRANCH}:
	\begin{algorithmic}[1]
		\STATE $\beta\gets 0$\,. $\triangleleft$ $\beta=0$ implies the left branch, 1 the right branch.
		\STATE Copy $n$, dimensionality of the parent hyperrectangle;
		\STATE Copy MIN, MAX, $\sum\limits_{t=0}^{n-1}\boldsymbol{x}\big(l(i_t)\big)$ , $\sum\limits_{t=0}^{n-1}\boldsymbol{x}\big(u(i_t)\big)$ , $l$ and $u$ from the parent.
		\STATE\label{contract}Update the copied parameters through contraction. If it fails, \textbf{return} a failure signal.
		\STATE\label{findt}$T\gets\{t|u(i_t)=l(i_t)\}$ , $n_z\gets|T|$ . $\triangleleft$ $|T|$ is the cardinality.
		\STATE $l\gets\{l(i_t)|t\notin T\}$ , $u\gets\{u(i_t)|t\notin T\}$ , $n\gets n-n_z$ .
		\STATE Update $\sum\limits_{t=0}^{n-1}\boldsymbol{x}\big(l(i_t)\big)$ and $\sum\limits_{t=0}^{n-1}\boldsymbol{x}\big(u(i_t)\big)$ .
		\STATE\label{push}Push $\{u(i_t)|t\in T\}$ in a global buffer $B$ that is to hold a qualified subset. $\triangleleft$ This step goes concurrently with Step \ref{findt} .
		\STATE $\kappa\gets\underset{t}{\argmin}\big(u(i_t)-l(i_t)\big)$ .
		\STATE $u^{\prime}\gets\{u(i_0),\ldots,u(i_\kappa)\}$ , $S(u)\gets\sum\limits_{t=0}^{n-1}\boldsymbol{x}\big(u(i_t)\big)$ .
		\STATE For $t\in[0,\,\kappa]$ , $u(i_t)\gets\min\big(u(i_t),\,\lfloor u(i_\kappa)/2\rfloor-\kappa+t\big)$ . $\triangleleft$ Loop $t$ from $\kappa$ and stop once $u(i_t)\leq\lfloor u(i_\kappa)/2\rfloor-\kappa+t$ .
		\STATE Update $\sum\limits_{t=0}^{n-1}\boldsymbol{x}\big(l(i_t)\big)$ , $\sum\limits_{t=0}^{n-1}\boldsymbol{x}\big(u(i_t)\big)$ . $\triangleleft$ Use $\mathcal{M}$ for fast update. 
	\end{algorithmic}
	If contraction succeeds in Step \ref{contract}, move to the right hyperrectangle in stack and execute the above steps again. Otherwise left-propagate through stack while erasing the last $n_z$ elements in buffer $B$ for each hyperrectangle, and stop once the current one has $\beta=0$ .
	
	\textbf{RIGHT BRANCH}:
	\begin{algorithmic}[1]
		\STATE $\beta\gets 1$ .
		\STATE For $t\in[0,\,\kappa]$ , $u(i_t)\gets u^{\prime}(i_t)$ ; $\sum\limits_{t=0}^{n-1}\boldsymbol{x}\big(u(i_t)\big)\gets S(u)$ .
		\STATE For $t\in[\kappa,\,n-1]$ , $l(i_t)\gets\max\big(l(i_t),\,u(i_\kappa)+1+t-\kappa\big)$ . $\triangleleft$ Loop $t$ from $\kappa$ and stop once $l(i_t)\geq u(i_\kappa)+1+t-\kappa$ .
		\STATE Update $\sum\limits_{t=0}^{n-1}\boldsymbol{x}\big(l(i_t)\big)$ . $\triangleleft$ Use $\mathcal{M}$ for fast update.
	\end{algorithmic}
	Move to the right hyperrectangle and execute \textbf{LEFT BRANCH}.
\end{algorithm}

\begin{figure}[t!]
	\includegraphics[width=\linewidth]{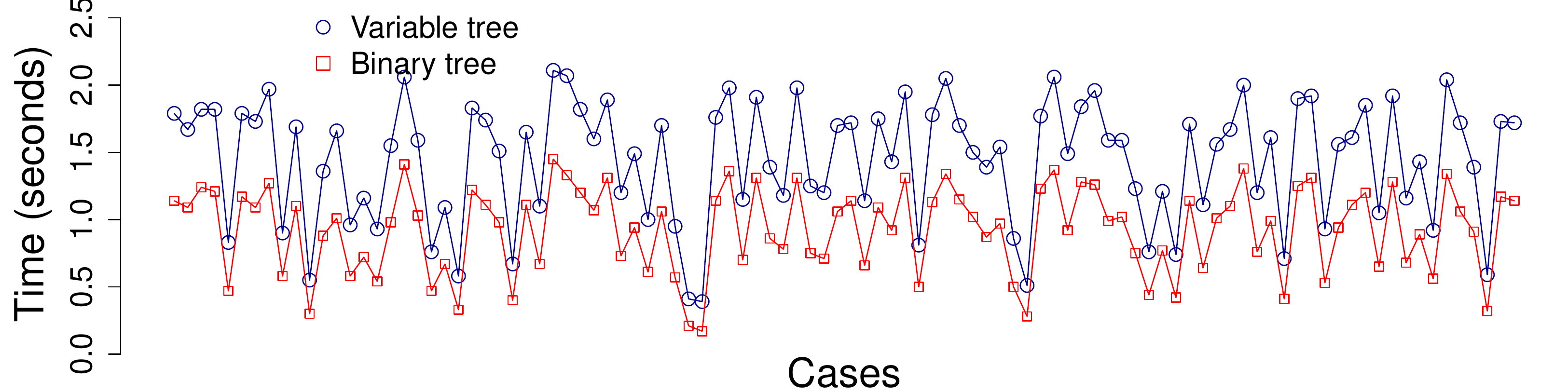}
	\caption{Supersets sizes = 70, dimensionality = 14, subsets sizes = 7, subset sum error thresholds = 0.01 for all dimensions, requested number of subsets $<\infty$, g++ '-O2' compile, 7 threads, Intel(R) i7-4770 CPU @ 3.40GHz, Windows 7. Preprocessing time included. Each of the 100 supersets contains random uniforms in [0, 10000]; the corresponding target sum is the sum of elements of a subset sampled at random. Binary subspcing yields about 1.6x acceleration.}\label{bvt}
\end{figure}

The current version of \pkg{FLSSS} differs in step (ii) and (iii): it halves the domain of the selected dimension, and reduces the dimensionality only if the domain width equals 1. Virtually, the current subspacing method constructs a binary tree while the previous ones construct a variable-branch tree. The binary tree appears to converge slower since it lengthens the path between each two nodes of dimension reduction, but it actually yields higher overall speed because (1) it could prune multiple branches at once if contraction fails over a halved hyperrectangle, and (2) child nodes will receive a smaller hyperrectangle (in terms of volume) which leads to faster contraction. Figure \ref{fg2} and Algorithm \ref{sbs} present details of subspacing. The speed benefit from binary subspacing will be more pronounced for multidimensional subset sum (Section \ref{mss}) where contractions consume the majority of mining time. 
%\mbox{Code \ref{binaryVariableTree}} 
The following code 
compares the time costs of mining a hundred supersets using the variable and binary subspacing methods. \mbox{Figure \ref{bvt}} shows the results.

%\begin{lstlisting}[caption={Compare computing speeds of binary and variable subspacing methods.}, label={binaryVariableTree}]
%set.seed(42)
%N = 70L; n = 7L; d = 14L
%mflsssBinTreeTime = numeric(100)
%mflsssVarTreeTime = numeric(100)
%for(i in 1L : 100L)
%{
%x = matrix(runif(N * d) * 10000, ncol = d)
%tmp = colSums(x[sample(1L : N, n), ])
%Sl = tmp - 0.01
%Su = tmp + 0.01
%rm(solution); gc()
%
%mflsssBinTreeTime[i] = system.time(FLSSS::mFLSSSpar(maxCore = 7, len = n, mV = x, mTarget = (Sl + Su) / 2, mME = (Su - Sl) / 2, solutionNeed = 1e3, tlimit = 3600))['elapsed']
%
%mflsssVarTreeTime[i] = system.time(FLSSS:::mFLSSSparVariableTree(maxCore = 7, len = n, mV = x, mTarget = (Sl + Su) / 2, mME = (Su - Sl) / 2, solutionNeed = 1e3, tlimit = 60))['elapsed']
%}
%\end{lstlisting}

\begin{CodeChunk}
\begin{CodeInput}
R> set.seed(42)
R> N = 70L; n = 7L; d = 14L
R> mflsssBinTreeTime = numeric(100)
R> mflsssVarTreeTime = numeric(100)
R> for(i in 1L : 100L)
+  {
+    x = matrix(runif(N * d, 0, 10000), ncol = d)
+    tmp = colSums(x[sample(1L : N, n), ])
+    Sl = tmp - 0.01
+    Su = tmp + 0.01
+    mflsssBinTreeTime[i] = system.time(FLSSS::mFLSSSpar(
+      maxCore = 7, len = n, mV = x, mTarget = (Sl + Su) / 2, 
+      mME = (Su - Sl) / 2, solutionNeed = 1e3, tlimit = 3600))['elapsed']
+    mflsssVarTreeTime[i] = system.time(FLSSS:::mFLSSSparVariableTree(
+      maxCore = 7, len = n, mV = x, mTarget = (Sl + Su) / 2, 
+      mME = (Su - Sl) / 2, solutionNeed = 1e3, tlimit = 3600))['elapsed']
R> }
R> mean(mflsssVarTreeTime / mflsssBinTreeTime)
\end{CodeInput}
\end{CodeChunk}
\begin{CodeChunk}
\begin{CodeOutput}
[1] 1.58791
\end{CodeOutput}
\end{CodeChunk}

\section{Multi-Subset Sum}\label{mss}
Given multiple sorted supersets and a subset size for each, the multi-Subset Sum seeks a subset from every superset such that elements in all subsets sum in a given range. The OFSSA directly applies to this problem following four steps: (i) shift elements in some or all supersets such that (ii) all elements in the shifted supersets constitute a nondecreasing sequence, a new superset; (iii) calculate a new subset sum target range in response to the shifting in (i); (iv) mine a subset of size equal to the sum of given subset sizes while the subset elements are bounded by the sub-supersets within the new superset.

Let $x_0,\ldots,x_{K-1}$ be $K$ sorted supersets of sizes $N_0,\ldots,N_{K-1}$. Let $n_0,\ldots,n_{K-1}$ be the respective subset sizes and $[\text{MIN}$, $\text{MAX}]$ be the target sum range. Shifting the supersets follows
\begin{equation}
x_h^\text{s}(t)\gets x_h(t) - x_h(0) + x_{h-1}(N_{K-1}-1),\,h\in[1,\,K),\,t\in[0,\,N_h)
\end{equation}
where $x_h^\text{s}$ denotes a shifted superset. Elements in shifted supersets are then pooled together as a new nondecreasing superset:
\begin{equation}
\begin{split}
x^\text{s}\gets\{\,&x^\text{s}_0(0),\,\ldots,\,x^\text{s}_0(N_0-1)\,,\\
&x^\text{s}_1(0),\,\ldots,\,x^\text{s}_1(N_1-1)\,,\\
&\hspace{5em}\vdots\\
&x^\text{s}_{K-1}(0),\,\ldots,\,x^\text{s}_{K-1}(N_{K-1}-1)\,\}\,.
\end{split}
\end{equation}
%$i(\sum_{h=0}^{h^\prime}n_h)$ to $i(\sum_{h=0}^{h^\prime+1}n_h)$
%A qualified subset is of size $\sum_{h=0}^{K-1}n_h$, and its $\sum_{h=0}^{sdfg}n_h$
%\begin{equation}
%x^\text{s}\gets\{x^\text{s}_0(0),\ldots,x^\text{s}_0(N_0-1),\,x^\text{s}_1(0),\ldots,x^\text{s}_1(N_1-1),\ldots,x^\text{s}_{K-1}(0),\ldots,x^\text{s}_{K-1}(N_{K-1}-1)\}\,.
%\end{equation}
The subset sum target range is adjusted by
\begin{equation}
\begin{split}
\text{MIN}^\text{s}&\gets\text{MIN} + \sum_{h=1}^{K-1}\big[x_{h-1}(N_{K-1}-1) - x_h(0)\big]\cdot n_h\,,\\
\text{MAX}^\text{s}&\gets\text{MIN}^\text{s}+(\text{MAX}-\text{MIN})\,.
\end{split}
\end{equation}
The multi-Subset sum seeks a monotonically increasing index array $i^\text{s}$ of size $\sum_{h=0}^{K-1}n_h$. Let $h^\prime\in[0,\,K)$. The initial hyperrectangle for contraction is outlined by
\begin{equation}
i^\text{s}(k)\in\Big[\sum_{h=0}^{h^\prime}N_h+k-\sum_{h=0}^{h^\prime}n_h,\,\sum_{h=0}^{h^\prime+1}N_h-n_{h^\prime}+k-\sum_{h=0}^{h^\prime}n_h\Big)\text{\,\,\,given\,\,\,}k\in\Big[\sum_{h=0}^{h^\prime}n_h,\,\sum_{h=0}^{h^\prime+1}n_h\Big)\,.
\end{equation}

\section{Multidimensional Subset Sum}\label{mdss}
A real superset of size $N$ in the $d$-dimensional space is an $N\times d$ matrix
\begin{equation}\label{mul}
\boldsymbol{x}=
\left(\begin{array}{ccc}
\boldsymbol{x}(0,0) & \ldots & \boldsymbol{x}(0,d-1)\\
\vdots & \vdots & \vdots\\
\boldsymbol{x}(N-1,0) & \ldots & \boldsymbol{x}(N-1,d-1)
\end{array}\right)=
\left(\begin{array}{c}
\boldsymbol{x}(0,)\\
\vdots\\
\boldsymbol{x}(N-1,)
\end{array}\right)=
\big(\boldsymbol{x}(,0),\,\ldots,\,\boldsymbol{x}(,d-1)\big)\;.
\end{equation}
Elementary algegrabs on two elements $\boldsymbol{x}(s,)$ and $\boldsymbol{x}(t,)$, $s,t\in[0,N)$, are defined element-wisely:
\begin{equation}
\begin{split}
&\boldsymbol{x}(s,)\pm\boldsymbol{x}(t,)=\big(\boldsymbol{x}(s,\,0)\pm\boldsymbol{x}(t,\,0),\,\ldots,\,\boldsymbol{x}(s,\,d-1)\pm\boldsymbol{x}(t,\,d-1)\big)\;,\\
&\boldsymbol{x}(s,)\leq\boldsymbol{x}(t,)\equiv\boldsymbol{x}(s,\,0)\leq\boldsymbol{x}(t,\,0)\land\ldots\land\boldsymbol{x}(s,\,d-1)\leq\boldsymbol{x}(t,\,d-1)\;.\\
\end{split}
\end{equation}
Given subset size $n$ and subset sum range [$S_L$, $S_U$] as two size-$d$ arrays, the multidimensional fixed-size Subset Sum algorithm (MFSSA) seeks an integer array $\boldsymbol{i}=(i_0,i_1,\ldots,i_{n-1})$ such that
\begin{equation}
S_L\leq\sum_{t=0}^{n-1}\boldsymbol{x}(i_t,)\leq S_U\;.
\end{equation}
The multidimensional variable-size Subset Sum follows similar conversion in Section \ref{odfsss}.

\subsection{Comonotonization}
If all columns in $\boldsymbol{x}$ are comonotonic \citep{tcocia}, $\boldsymbol{x}$ can be sorted so that $\boldsymbol{x}(0,)\leq\ldots\leq\boldsymbol{x}(N-1,)$ . Overloading arithmetic operators \citep{ooai} in OFSSA then solves the problem.

In general, MFSSA roughly consists of (i) padding an extra column of non-decreasing integers to the superset, (ii) scaling and adding this column to the rest to make all columns comonotonic (\textit{comonotonize} / \textit{comonotonization}), and (iii) mining for at most $N(N-n)/2+1$ subset sum ranges regarding the new superset. Each of these subset sum ranges corresponds to a subset sum in the extra column. Mining different subset sum ranges in (iii) are independent and share the same auxiliary matrix of $\mathcal{M}$ (\mbox{Section \ref{qtm}}), thus can employ multithreading.

Let $\boldsymbol{x}^*$ be the comonotonized superset. The extra column $\boldsymbol{x}^*(,d)$ is referred to as the \textit{key column}. For convenience, let array $v$ refer to $\boldsymbol{x}^{*}(,d)$ . The key column is constructed by
\begin{equation}\label{eq17}
\begin{split}
v(0)&\gets 0\;,\text{ and for }s\in[1,\,N)\;:\\
v(s)&\gets v(s-1)\textbf{\, If \,}\boldsymbol{x}(s-1,)\leq\boldsymbol{x}(s,)\;,\\
v(s)&\gets v(s-1)+1\textbf{ Otherwise}\,.
\end{split}
\end{equation}
%\begin{equation}
%\begin{split}
%\boldsymbol{x}^{*}(0,\,d)&\gets 0\;,\text{ and for }s\in[1,\,N-1]\;:\\
%\boldsymbol{x}^{*}(s,\,d)&\gets\boldsymbol{x}^{*}(s-1,\,d)\textbf{\, If \,}\boldsymbol{x}(s-1,)\leq\boldsymbol{x}(s,)\;,\\
%\boldsymbol{x}^{*}(s,\,d)&\gets\boldsymbol{x}^{*}(s-1,\,d)+1\textbf{ Otherwise.}
%\end{split}
%\end{equation}
Let $\Delta\boldsymbol{x}(,t)$ be the discrete differential of $\boldsymbol{x}(,t)$, $t\in[0,\,d)$. The rest columns of $\boldsymbol{x}^{*}$ are computed by
\begin{equation}\label{eq18}
\begin{split}
&\theta(t)=\Big|\min\Big(0,\,\min\big(\Delta\boldsymbol{x}(,t)\big)\Big)\Big|\text{ and}\\
&\boldsymbol{x}^{*}(,t)\gets\boldsymbol{x}(,t)+v\cdot\theta(t)\;,
\end{split}
\end{equation}
where $\theta(t)$ is referred to as the \textit{comonotonization multiplier} for $\boldsymbol{x}(,t)$. The key column $v$ has no subset sum constraint. However, because it is a sorted integer sequence with maximal discrete differential of 1, all unique subset sums in $v$ compose an integer sequence:
\begin{equation}
S^\text{key}=\Big(\sum_{t=0}^{n-1}v(t),\,1+\sum_{t=0}^{n-1}v(t),\,2+\sum_{t=0}^{n-1}v(t),\,\ldots,\,\sum_{t=N-n}^{N-1}v(t)\Big)\,.
\end{equation}
%\begin{equation}
%S^\text{key}=\Big(\sum_{t=0}^{n-1}\boldsymbol{x^*}(t,\,d),\,1+\sum_{t=0}^{n-1}\boldsymbol{x^*}(t,\,d),\ldots,\,\sum_{t=N-n}^{N-1}\boldsymbol{x^*}(t,\,d)\Big)\,.
%\end{equation}
The size of $S^\text{key}$ equals $\sum_{t=N-n}^{N-1}v(t)-\sum_{t=0}^{n-1}v(t)+1$, which would be no more than $N(N-n)/2+1$. Let $N_S$ be the size of $S^\text{key}$. 
%Every element in $S^\text{key}$ yields a new subset sum range to mine. 
We have the following $N_S$ subset sum ranges to mine:
\begin{equation}\label{eq22}
\begin{split}
\boldsymbol{S}^*_L&=\left[\begin{array}{cccc}
S^\text{key}(0)\theta(0)+S_L(0) & \ldots & S^\text{key}(0)\theta(d-1)+S_L(d-1) & S^\text{key}(0)\\
\vdots & \vdots & \vdots & \vdots\\
S^\text{key}(N_S-1)\theta(0)+S_L(0) & \ldots & S^\text{key}(N_S-1)\theta(d-1)+S_L(d-1) & S^\text{key}(N_S-1)
\end{array}\right]\\
%\end{equation}
%\begin{equation}
\boldsymbol{S}^*_U&=\left[\begin{array}{cccc}
S^\text{key}(0)\theta(0)+S_U(0) & \ldots & S^\text{key}(0)\theta(d-1)+S_U(d-1) & S^\text{key}(0)\\
\vdots & \vdots & \vdots & \vdots\\
S^\text{key}(N_S-1)\theta(0)+S_U(0) & \ldots & S^\text{key}(N_S-1)\theta(d-1)+S_U(d-1) & S^\text{key}(N_S-1)
\end{array}\right]
\end{split}
\end{equation}
where [$\boldsymbol{S}^*_L(s,)$, $\boldsymbol{S}^*_U(s,)$], $s\in[0,\,N_S)$, account for one subset sum range.

Consider the following toy example of finding a size-2 subset from a 2D superset ($n=2,\,d=2,\,N=3$)
\begin{equation}
\boldsymbol{x}=
\left(\begin{array}{cc}
4 & 10\\
2 & 25\\
8 & 17
\end{array}\right)
\end{equation}
with $S_L=(11,\,26)$ and $S_U=(12,\,28)$. The minimal discrete differentials of the two columns are -2 and -8. Then the comonotonization multipliers are 2 and 8 respectively. We comonotonize $\boldsymbol{x}$ by
\begin{equation}\label{mflsssexample}
\boldsymbol{x}^{*}=
\left(\begin{array}{ccc}
4 + 0\times 2 & 10 + 0\times 8 & 0\\
2 + 1\times 2 & 25 + 1\times 8 & 1\\
8 + 2\times 2 & 17 + 2\times 8 & 2
\end{array}\right)
=\left(\begin{array}{ccc}
4 & 10&0\\
4 & 33&1\\
12 & 33&2
\end{array}\right)
\end{equation}
according to Equations \eqref{eq17}, \eqref{eq18}. The unique size-two subset sums in the key column are 1, 2, 3, thus
\begin{equation}
\begin{split}
\boldsymbol{S}^*_L&=\left[\begin{array}{ccc}
11+1\times 2 & 26+1\times 8 & 1\\
11+2\times 2 & 26+2\times 8 & 2\\
11+3\times 2 & 26+3\times 8 & 3\\
\end{array}\right]=\left[\begin{array}{ccc}
13 & 34 & 1\\
15 & 42 & 2\\
17 & 50 & 3
\end{array}\right]\,,\\
\boldsymbol{S}^*_U&=\left[\begin{array}{ccc}
12+1\times 2 & 28+1\times 8 & 1\\
12+2\times 2 & 28+2\times 8 & 2\\
12+3\times 2 & 28+3\times 8 & 3\\
\end{array}\right]=\left[\begin{array}{ccc}
14 & 36 & 1\\
16 & 44 & 2\\
18 & 52 & 3
\end{array}\right]\,.
\end{split}
\end{equation}
For $\boldsymbol{x}^{*}$, there are 3 subset sum ranges subject to mining.

\subsection{Order optimizations}\label{op}
The order optimizations have two folds: (a) reordering rows of the superset before comonotonization and (b) reordering the subset sum ranges. Both can accelerate subset mining. %The conclusions are mainly based on observations from simulations rather than theoretical inferences. 
Several conjectures for the speedup are given at the end of the section.

Before adding the key column, we sort rows of $\boldsymbol{x}$ by one of its columns, the \textit{leader column}, in ascending order. The leader column has the least sum of squares of rank differences, or, the greatest sum of Spearman's correlations \citep{tpamo}, with other columns. In a sense, the leader column \textit{correlates the rest columns the most}. Comparing sums of Spearman's correlations is not the only way to define a leader column. We choose the method because it is most common and computationally cheap.

The next optimization reorders the rows of $\boldsymbol{S}_L^*$ and $\boldsymbol{S}_U^*$ based on their likelihoods of yielding qualified subsets.  
%estimates which row of $\boldsymbol{S}_L^*$ and $\boldsymbol{S}_U^*$ would more likely yield qualified subsets, and then switches these rows  
Let $t^\prime$ be the index of the leader column. After sorting $\boldsymbol{x}$ by $\boldsymbol{x}(,t^\prime)$, we estimate the percentile of the leader column's subset sum target within the range of all possible subset sums:
\begin{equation}
p=\frac{[S_L(,t^\prime)+S_U(,t^\prime)]\,/\,2-\sum_{s=0}^{n-1}\boldsymbol{x}(s,\,t^\prime)}{\sum_{s=N-n}^{N-1}\boldsymbol{x}(s,\,t^\prime)-\sum_{s=0}^{n-1}\boldsymbol{x}(s,\,t^\prime)}\,.
\end{equation}
If there exist qualified subsets, their subset sums in the \textit{key column} should have percentiles close to $p$. We prioritize the subset sum targets whose percentiles are close to $p$, thus the rows of $\boldsymbol{S}^*_L$ and $\boldsymbol{S}^*_U$ are ordered by
\begin{equation}
\Big|\frac{\boldsymbol{S}^*_U(,d)-\boldsymbol{S}^*_U(0,\,d)}{\boldsymbol{S}^*_U(N-1,\,d)-\boldsymbol{S}^*_U(0,\,d)}-p\Big|\,.
\end{equation}

Given $m$ threads, the first $m$ rows of $\boldsymbol{S}_L^*$ and $\boldsymbol{S}_U^*$ are mined concurrently. The first finished thread then works on the $m+1$ th rows so forth if the current number of qualified subsets is unsatisfied. The threads are scheduled by several atomic class objects \citep{itbb}. The scheduling overhead is negligible. 
\begin{figure}[t!]
	\centering
	\includegraphics[width=\textwidth]{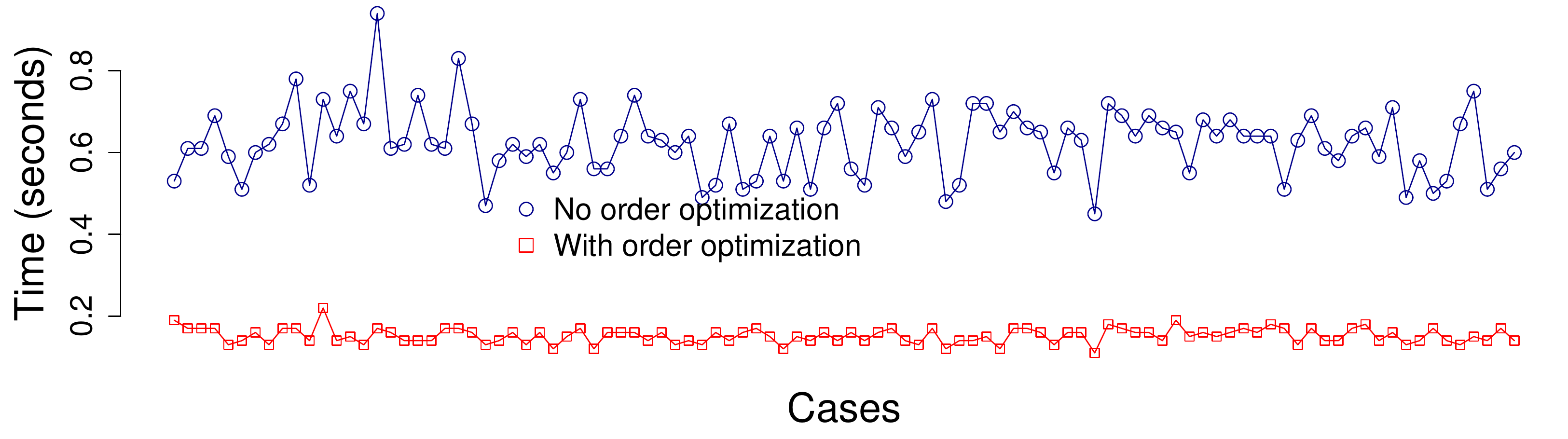}
	\caption{A hundred $60\times5$ supersets and subset sum targets generated at random, subset size $n=6$, g++ '-O2' compile, 7 threads, Intel(R) i7-4770 CPU @ 3.40GHz, Windows 7. Preprocessing time included. Order optimization yields about 4x acceleration for finding all qualified subsets.}\label{orderopt}
\end{figure}

%\renewcommand{\lstlistingname}{Code}
%\begin{lstlisting}[caption={Order optimization benchmarking.},label={bswowoo}]
%set.seed(42)
%N = 60L; n = 6L; d = 5L
%noOpt = numeric(100)   # For keeping the time costs of 100 test cases
%withOpt = numeric(100) # with and without order optimization.
%for(i in 1L : 100L)
%{
%x = matrix(runif(N * d) * 10000, ncol = d)
%solution = sample(1L : N, n)
%Sl = colSums(x[solution, ]) * 0.999 # Subset sum range lower bound.
%Su = Sl / 0.999 * 1.001             # Upper bound.
%rm(solution); gc()
%
%noOpt[i] = system.time(FLSSS::mFLSSSparImposeBounds(maxCore = 7, len = n, mV = x, mTarget = (Sl + Su) / 2, mME = (Su - Sl) / 2, solutionNeed = 1e6))['elapsed'] 
%# It is extremely unlikely that the number of qualified subsets for such a small subset sum range exceeds 1 million.
%
%withOpt[i] = system.time(FLSSS::mFLSSSpar(maxCore = 7, len = n, mV = x, mTarget = (Sl + Su) / 2, mME = (Su - Sl) / 2, solutionNeed = 1e6))['elapsed']
%}
%\end{lstlisting}

%Code \ref{bswowoo} 
The following code 
demonstrates the speed advantage from order optimizations in a hundred test cases. Figure \ref{orderopt} shows the results.
\begin{CodeChunk}
\begin{CodeInput}
R> set.seed(42)
R> N = 60L; n = 6L; d = 5L
R> noOpt = numeric(100) 
R> withOpt = numeric(100)
R> for(i in 1L : 100L)
+  {
+    x = matrix(runif(N * d) * 10000, ncol = d)
+    solution = sample(1L : N, n)
+    Sl = colSums(x[solution, ]) * 0.999
+    Su = Sl / 0.999 * 1.001
+    rm(solution); gc()	
+    noOpt[i] = system.time(FLSSS::mFLSSSparImposeBounds(maxCore = 7, 
+      len = n, mV = x, mTarget = (Sl + Su) / 2, mME = (Su - Sl) / 2, 
+      solutionNeed = 1e6))['elapsed']
+    withOpt[i] = system.time(FLSSS::mFLSSSpar(maxCore = 7, len = n, 
+      mV = x, mTarget = (Sl + Su) / 2, mME = (Su - Sl) / 2, 
+      solutionNeed = 1e6))['elapsed']
+  }
R> mean(noOpt / withOpt)
\end{CodeInput}
\end{CodeChunk}
\begin{CodeChunk}
\begin{CodeOutput}
[1] 4.392401
\end{CodeOutput}
\end{CodeChunk}
These simulations seek all qualified subsets, thus order optimization (b) has no effect, because all rows in $\boldsymbol{S}_L^*$ and $\boldsymbol{S}_U^*$ will be in trial. 
%Accelerations shown in Figure \ref{orderopt} purely come from superset reordering. 
A certain yet minor reason for the acceleration due to superset reordering is that it can lower the number of unique elements in the key column by \mbox{Equation \eqref{eq17}}, thus leads to fewer rows in $\boldsymbol{S}_L^*$ and $\boldsymbol{S}_U^*$ and fewer tasks for the computing threads. A probable and major reason is that reordering the superset puts its elements in compact shapes or clusters instead of random, scattered formations in multidimensional space, which leads to (i) more intense hyperrectangle contraction (\mbox{Section \ref{Contraction}}) navigated by those compact shapes or clusters, and thus (ii) fewer child hyperrectangles spawned for predicting the locations of qualified subsets.

\subsection{Integer compression}\label{icp}
To further accelerate the mining speed with less concern about the accuracy, we round superset $\boldsymbol{x}$ %(Equation \eqref{mul}),
as $\boldsymbol{x}^z$, and then compress every row of the comonotonized superset $\boldsymbol{x}^{z*}$ 
%(Equation \eqref{eq18}) 
into an array of 64-bit integer buffers. The final superset is denoted by $\boldsymbol{x}^{z*c}$. The transformation from $\boldsymbol{x}$ to $\boldsymbol{x}^{z*c}$ is referred to as \textit{integerization}. The consequent dimension reduction enhances cache locality \citep{tlp} and thus computing speed.

In MFSSA, shifting and scaling a column and its subset sum range does not affect mining results. Let $\lambda(t)$, an integer, be the user-defined maximum of column $\boldsymbol{x}^{z}(,t)$. We shift, scale and round a column of $\boldsymbol{x}$ and the corresponding subset sum range by
\begin{equation}\label{inte}
\begin{split}
\boldsymbol{x}^{z}(,t)&\gets\bigg\lfloor\frac{\boldsymbol{x}(,t)-\min\big(\boldsymbol{x}(,t)\big)}{\max\big(\boldsymbol{x}(,t)\big)}\cdot\lambda(t)\bigg\rceil\,,\\
S_L^z(t)&\gets\bigg\lfloor\frac{S_L(t)-\min\big(\boldsymbol{x}(,t)\big)\cdot n}{\max\big(\boldsymbol{x}(,t)\big)}\cdot\lambda(t)\bigg\rceil\,,\\
S_U^z(t)&\gets\bigg\lfloor\frac{S_U(t)-\min\big(\boldsymbol{x}(,t)\big)\cdot n}{\max\big(\boldsymbol{x}(,t)\big)}\cdot\lambda(t)\bigg\rceil\,.\\
\end{split}
\end{equation}
The above equations guarantee nonnegative integers in $\boldsymbol{x}^{z}(,t)$ with minimum 0 and maximum $\lambda(t)$. Without considering numeric errors brought by scaling and shifting, larger $\lambda(t)$ makes $\boldsymbol{x}^{z}$ and $\boldsymbol{x}$ have closer joint distributions, and the chance of the two yielding different qualified subsets (element index-wise) will be lower.

The comonotonization of $\boldsymbol{x}^z$ results in $\boldsymbol{x}^{z*}$,   $\boldsymbol{S}^{z*}_L$ and $\boldsymbol{S}^{z*}_U$. Compressing integers for dimension reduction approximately consists of (i) finding the \textit{largest absolute value that could be reached during mining} (LVM) for each dimension (column) of $\boldsymbol{x}^{z*}$, (ii) calculating how many bits are needed to represent the LVM for each dimension and how many 64-bit buffers are needed to store those bits for all dimensions, (iii) cramming each row of $\boldsymbol{x}^{z*}$ into the buffers and (iv) defining compressed integer algebras through bit-manipulation.

The LVM for a certain dimension should equal or exceed the maximum of the absolute values of all temporary or permanent variables in the mining program within that dimension. Let $\psi(t)$ be the LVM for $\boldsymbol{x}^{z*}(t)$, $t\in[0,\,d)$. We have
\begin{equation}
\psi(t)\gets\max\Big(\max\big(\big|\boldsymbol{S}^{z*}_L(,t)\big|\big),\,\max\big(\big|\boldsymbol{S}^{z*}_U(,t)\big|\big), \sum_{s=N-n}^{N-1}\boldsymbol{x}^{z*}(s,\,t)\Big)\,,
\end{equation}
and the number of bits alloted to dimension $t$ equals
\begin{equation}
\beta(t)\gets\big\lceil\log_2\big(\psi(t)\big)\big\rceil+1\,.
\end{equation}
The extra bit is used as the sign bit for comparison. 

For the first 64-bit integer in the $s$ th row of $\boldsymbol{x}^{z*c}$, $s\in[0,\,N)$,
\begin{equation}\label{sa}
\begin{split}
\boldsymbol{x}^{z*c}(s,\,0)\gets&\boldsymbol{x}^{z*}(s,\,0)\ll\big(64-\beta(0)\big)\,+\\
&\boldsymbol{x}^{z*}(s,\,1)\ll\big(64-\beta(0)-\beta(1)\big)\,+\\
&\boldsymbol{x}^{z*}(s,\,2)\ll\big(64-\beta(0)-\beta(1)-\beta(2)\big)\,+\\
&\hspace{5em}\vdots
\end{split}
\end{equation}
where $\ll$ is the left bit-shift operator. The term $64-\beta(0)$ or $64-\beta(0)-\beta(1)$ $\ldots$ is referred to as shift distance. The above construction for $\boldsymbol{x}^{z*c}(s,\,0)$ stops once the shift distance would become negative. The next 64-bit buffer $\boldsymbol{x}^{z*c}(s,\,1)$ then starts accommodating bits of the rest elements in $\boldsymbol{x}^{z*}(s,)$, so on and so forth. 

Let $d^c$ be the dimensionality of $\boldsymbol{x}^{z*c}$. It is estimated before integer compression.
A different order of $\boldsymbol{x}^{z*}$ columns means a different order of $\beta$, and such order may lead to a smaller $d^c$. Finding the best order for minimizing $d^c$ accounts for a bin-packing problem \citep{bpacking}. Currently \pkg{FLSSS} does not optimize this order for giving users the option of bounding partial columns (see package user manual), which needs to put the lower-bounded (upper-bounded) columns next to each other. Equation \eqref{sa} also applies to computing $\boldsymbol{S}_L^{z*c}$ and $\boldsymbol{S}_U^{z*c}$.

\subsection{Comparison mask}
The comparison mask is responsible for comparing two compressed integer arrays. Similar to \mbox{Equation \eqref{sa}}, a 64-bit buffer array $\pi$ of size $d^c$ is constructed by
\begin{equation}
\begin{split}
\pi(0)\gets&1\ll\big(64-\beta(0)\big) +\\
&1\ll\big(64-\beta(0)-\beta(1)\big) +\\
&\hspace{5em}\vdots
\end{split}
\end{equation}
In $\pi$, bits equal to 1 align with the sign bits in a row of $\boldsymbol{x}^{z*c}$. We define
\begin{equation}\label{addsubdef}
\begin{split}
&\boldsymbol{x}^{z*c}(s,)\pm\boldsymbol{x}^{z*c}(t,)=\big(\boldsymbol{x}^{z*c}(s,\,0)\pm\boldsymbol{x}^{z*c}(t,\,0),\,\ldots,\,\boldsymbol{x}^{z*c}(s,\,d^c-1)\pm\boldsymbol{x}^{z*c}(t,\,d^c-1)\big)\,,\\
%&\boldsymbol{x}^{z*c}(s,)\leq\boldsymbol{x}^{z*c}(t,)\equiv\boldsymbol{x}^{z*c}(s,\,0)\leq\boldsymbol{x}^{z*c}(t,\,0)\land\ldots\land\boldsymbol{x}^{z*c}(s,\,d^c-1)\leq\boldsymbol{x}^{z*c}(t,\,d^c-1)\,.\\
&\boldsymbol{x}^{z*c}(s,)\leq\boldsymbol{x}^{z*c}(t,)\equiv\big[\big(\boldsymbol{x}^{z*c}(t,\,0)-\boldsymbol{x}^{z*c}(s,\,0)\big)\,\&\,\pi(0)=0\big]\,\land\\
&\hspace{9.7em}\big[\big(\boldsymbol{x}^{z*c}(t,\,1)-\boldsymbol{x}^{z*c}(s,\,1)\big)\,\&\,\pi(1)=0\big]\,\land\\
&\hspace{20em}\vdots\\
&\hspace{9.7em}\big[\big(\boldsymbol{x}^{z*c}(t,\,d^c-1)-\boldsymbol{x}^{z*c}(s,\,d^c-1)\big)\,\&\,\pi(d^c-1)=0\big]
\end{split}
\end{equation}
where \& is the bitwise AND operator. 

\begin{figure}[t!]
	\includegraphics[width=\linewidth]{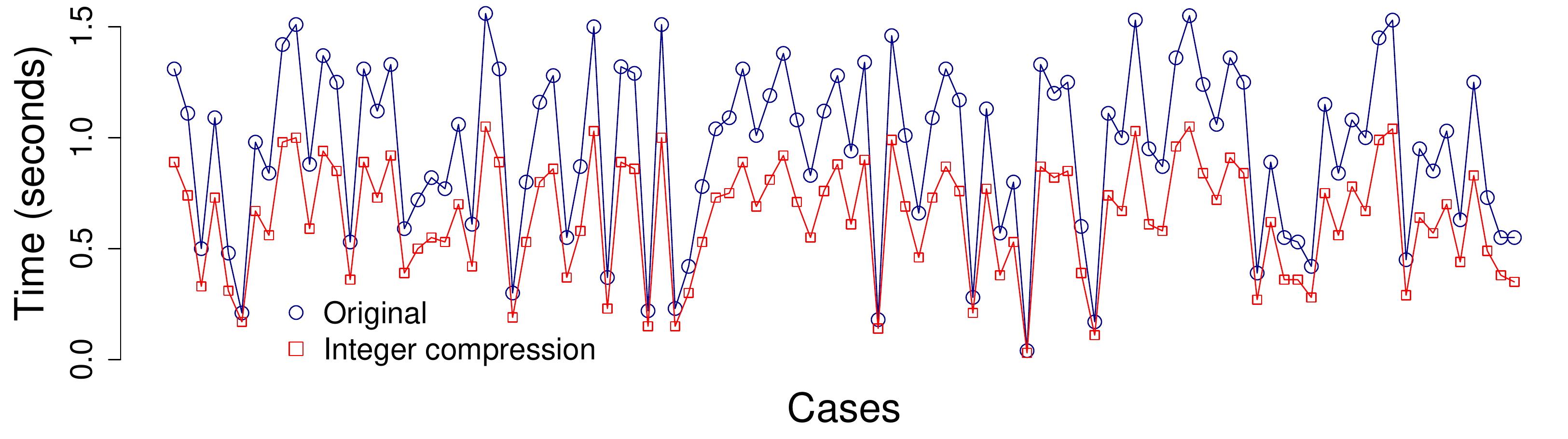}
	\caption{A hundred $70\times14$ supersets and subset sum targets generated at random, subset size $n=7$, g++ '-O2' compile, 7 threads, Intel(R) i7-4770 CPU @ 3.40GHz, Windows 7. Preprocessing time included. Integerization yields about 1.5x acceleration for finding all qualified subsets.}\label{compressopt}
\end{figure}

Consider the following toy example where $d^c=1$. Five elemental integers constitute one 64-bit buffer, and the comparison mask is
\begin{equation*}
\texttt{comparisonMask}=\boldsymbol{1}000000000\boldsymbol{1}0000\boldsymbol{1}000000000\boldsymbol{1}000000000000\boldsymbol{1}0000000000000000000000000\,.
\end{equation*}
The mask indicates the first elemental integer occupies the 2nd to 9th bits, the second elemental integer occupies the 12th to 15th bits, and so on. The following C++ codes define addition, subtraction and comparison of such two 64-bit buffers on a 64-bit machine:
\begin{verbatim}
std::size_t add(std::size_t x, std::size_t y) { return x + y; }

std::size_t subtract(std::size_t x, std::size_t y) { return x - y; }

bool lessEqual(std::size_t x, std::size_t y, std::size_t comparisonMask)
{ return (y - x) & signMask == 0; }
\end{verbatim}
The addition and subtraction have no overheads comparing to those for normal integers. The comparison may halve the speed of a single `$\leq$' operation, but acceleration due to dimension reduction would reverse the situation globally. See \citep{asacci} for more discussions on compressed integer algebras. 
% Code \ref{compresscode} and 
The following code demonstrates the speedup from integerization. Figure \ref{compressopt} shows the results.

%\renewcommand{\lstlistingname}{Code}
%\begin{lstlisting}[caption={Benchmark mining speeds with and without integerization.}, label={compresscode}]
%N = 70L; n = 7L; d = 14L
%noOpt = numeric(100)
%withOpt = numeric(100)
%for(i in 1L : 100L)
%{
%x = matrix(runif(N * d) * 10000, ncol = d)
%solution = sample(1L : N, n)
%Sl = colSums(x[solution, ]) * 0.999
%Su = Sl / 0.999 * 1.001
%rm(solution); gc()
%
%noOpt[i] = system.time(FLSSS::mFLSSSpar(maxCore = 7, len = n, mV = x, mTarget = (Sl + Su) / 2, mME = (Su - Sl) / 2, solutionNeed = 1e3, tlimit = 3600))['elapsed']
%noOpt[i] = noOptRst[[i]]$tc
%
%# Round the superset and compress integers:
%withOpt[i] = system.time(FLSSS::mFLSSSparIntegerized(maxCore = 7, len = n, mV = x, mTarget = (Sl + Su) / 2, mME = (Su - Sl) / 2, solutionNeed = 1e3, tlimit = 3600))['elapsed']
%}
%\end{lstlisting}

\begin{CodeChunk}
\begin{CodeInput}
R> set.seed(42)
R> N = 70L; n = 7L; d = 14L
R> noOpt = numeric(100)
R> withOpt = numeric(100)
R> for(i in 1L : 100L)
+  {
+    x = matrix(runif(N * d) * 10000, ncol = d)
+    solution = sample(1L : N, n)
+    Sl = colSums(x[solution, ]) * 0.999
+    Su = Sl / 0.999 * 1.001
+    rm(solution); gc()	
+    noOpt[i] = system.time(FLSSS::mFLSSSpar(maxCore = 7, len = n, mV = x, 
+      mTarget = (Sl + Su) / 2, mME = (Su - Sl) / 2, solutionNeed = 1e3, 
+      tlimit = 3600))['elapsed']
+    withOpt[i] = system.time(FLSSS::mFLSSSparIntegerized(maxCore = 7, 
+      len = n, mV = x, mTarget = (Sl + Su) / 2, mME = (Su - Sl) / 2, 
+      solutionNeed = 1e3, tlimit = 3600))['elapsed']
+  }
R> mean(noOpt / withOpt)
\end{CodeInput}
\end{CodeChunk}
\begin{CodeChunk}
\begin{CodeOutput}
[1] 1.477555
\end{CodeOutput}
\end{CodeChunk}

\section{Multidimensional Knapsack problem}\label{mkp}
The Knapsack problem, especially the 0-1 Knapsack problem, is one of the oldest combinatorial optimization problems and has been extensively studied \citep{kprob}. Given a set of items with a profit attribute and a cost attribute, the 0-1 Knapsack problem seeks a subset of items to maximize the total profit while the total cost does not surpass a given value. The multidimensional 0-1 Knapsack problem assigns multiple cost attributes to an item, and maximizes the total profit while the total cost in each cost dimension stays below their individual upper bounds. The computational complexity of the multidimensional 0-1 Knapsack problem escalates rapidly as the dimensionality rises.

MFSSA directly applies to the multidimensional fixed-size 0-1 Knapsack (MF01K) problem. See \mbox{Section \ref{odfsss}} for converting a variable-size instance. Consider an MF01K instance with subset size $n$, cost attributes $d$, items $N$. The costs constitute an $N\times d$ superset $\boldsymbol{x}$. Rows of $\boldsymbol{\boldsymbol{x}}$ are sorted by item profits in ascending order. For $t\in[0,\,d)$, the subset sum upper bound $S_U(t)$ equals the given cost upper bound, and the lower bound $S_L(t)$ equals the sum of the least $n$ elements in $\boldsymbol{x}(t)$. We then pad the key column in Equation \eqref{eq17}, and comonotonize $\boldsymbol{x}$ to obtain $\boldsymbol{x}^*$ in Equation \eqref{eq18} and $\boldsymbol{S}^*_L$, $\boldsymbol{S}^*_U$ in Equation \eqref{eq22}.

\subsection{Optimization}\label{kopt}
Column $\boldsymbol{S}^*_L(,d)$ or $\boldsymbol{S}^*_U(,d)$ essentially consists of sums of ranks of the item profits, thus a qualified subset found via mining [$\boldsymbol{S}^*_L(s,),\,\boldsymbol{S}^*_U(s,)$], $s\in[1,\,N)$ would more likely have a greater total item profit than that from [$\boldsymbol{S}^*_L(s^\prime,),\,\boldsymbol{S}^*_U(s^\prime,)$], $s>s^\prime$. Therefore, we sort $\boldsymbol{S}^*_L$ and $\boldsymbol{S}^*_U$ by $\boldsymbol{S}^*_U(,d)$ in descending order to increase the chance of qualified subsets having higher total item profits being found sooner. On the other hand, a heuristic approach stops immediately once it finds a qualified subset.

Given $m$ threads, instead of mining the first $m$ rows of $\boldsymbol{S}^*_L$ and $\boldsymbol{S}^*_U$ concurrently (\mbox{Section \ref{op}}), the threads all concentrate on [$\boldsymbol{S}^*_L(0,),\,\boldsymbol{S}^*_U(0,)$] first. Given a constant $\phi$, we perform a breadth-first search starting from the root node of the binary tree in Figure \ref{fg2} until we have no less than $m\phi$ nodes for trail in the same hierarchy. The threads then have $m\phi$ independent tasks to work on. These tasks have heterogeneous difficulties. To lower the chance of thread idleness, the first $m$ tasks are solved concurrently, and the first finished thread moves onto the $m+1$ th task and so forth. If idle threads are detected, another breadth-first expansion generates sufficient tasks to keep them busy. 

Once a thread finds a qualified subset, it updates the current optimum if the total profit of the subset exceeds that of the current optimum. The optimum and its profit are guarded by a spin mutex lock \citep{mutexlock}. 

%We exam the total item profit of the upper bounds of any hyper
%For any hyperrectangle yielded from contraction, we exam 

After each contraction (Section \ref{Contraction}), if the total item profit of the hyperrectangle's upper bounds $u$ is below that of the current optimum, we prune the entire branch rooted at this hyperrectangle.

\section{Generalized Assignment problem}\label{gap}
The Generalized Assignment Problem (GAP) is an NP-hard combinatorial optimization problem \citep{gaproblem}. It assigns $T$ tasks to $A$ agents where each agent can take zero to all tasks. A task would both profit and cost an agent, and every agent has a budget. The GAP seeks an assignment to maximize the total profit.

Let $\boldsymbol{c}$ and $\boldsymbol{p}$ be the cost and profit matrix:
\begin{equation}
%\begin{split}
\boldsymbol{c}=\left[\begin{array}{ccc}
\boldsymbol{c}(0,\,0) & \ldots & \boldsymbol{c}(0,\,A-1)\\
\vdots & \vdots & \vdots\\
\boldsymbol{c}(T-1,\,0) & \ldots & \boldsymbol{c}(T-1,\,A-1)
\end{array}\right],\;
\boldsymbol{p}=\left[\begin{array}{ccc}
\boldsymbol{p}(0,\,0) & \ldots & \boldsymbol{p}(0,\,A-1)\\
\vdots & \vdots & \vdots\\
\boldsymbol{p}(T-1,\,0) & \ldots & \boldsymbol{p}(T-1,\,A-1)
\end{array}\right]
\end{equation}
where $\boldsymbol{c}(s,\,t)$ and $\boldsymbol{p}(s,\,t)$ are the cost and profit from assigning task $s$ to agent $t$. We integrate $\boldsymbol{c}$ and $\boldsymbol{p}$ to a multidimensional superset:
\begin{equation}\label{integrate}
\boldsymbol{x}=\left[\begin{array}{cccc|c}
\boldsymbol{c}(0,\,0) & 0 & \ldots & 0 & \boldsymbol{p}(0,\,0)\\
0 & \boldsymbol{c}(0,\,1) & \ldots & 0 & \boldsymbol{p}(0,\,1)\\
\vdots & \vdots & \ddots & \vdots & \vdots\\
0 & 0 & 0 & \boldsymbol{c}(0,\,A-1) & \boldsymbol{p}(0,\,A-1)\\
\cline{1-5}
\boldsymbol{c}(1,\,0) & 0 & \ldots & 0 & \boldsymbol{p}(1,\,0)\\
0 & \boldsymbol{c}(1,\,1) & \ldots & 0 & \boldsymbol{p}(1,\,1)\\
\vdots & \vdots & \ddots & \vdots & \vdots\\
0 & 0 & 0 & \boldsymbol{c}(1,\,A-1) & \boldsymbol{p}(1,\,A-1)\\
\cline{1-5}
\vdots & \vdots & \ddots & \vdots & \vdots\\
0 & 0 & 0 & \boldsymbol{c}(T-1,\,A-1) & \boldsymbol{p}(T-1,\,A-1)
\end{array}\right]\,.
\end{equation}
This $(T\times A)\times (A+1)$ superset has $T$ \textit{blocks} of $A$ rows. Each block is transformed from a row in $\boldsymbol{c}$ and in $\boldsymbol{p}$. For instance, the first $A$ rows of $\boldsymbol{x}$ constitute the first block where the diagonal entries equal $\boldsymbol{c}(0,)$ and the last column equals $\boldsymbol{p}(0,)$. 

Given agent budgets $\boldsymbol{b}=\big(\boldsymbol{b}(0),\ldots,\boldsymbol{b}(A-1)\big)$, the GAP is equivalent to a multidimensional subset sum problem of finding a subset of size $T$ from $\boldsymbol{x}$ subject to: (i) the subset sums of the first $A$ columns are no greater than $\boldsymbol{b}$, (ii) each of the $T$ blocks contributes one element to the subset, and (iii) the subset sum is maximized for the profit column.

Within each block, we sort the rows by the profit column $\boldsymbol{x}(,A)$ in ascending order. Then $\boldsymbol{x}(,A)$ is replaced with the key column according to Equation \eqref{eq17}, block by block. 
The aforementioned constraint (ii) changes the initial hypercube (Equation \eqref{ineq2}) to
\begin{equation}
i_k\in[Ak,\,Ak+A),\;k\in\{0,1,\ldots,T-1\}\;.
\end{equation}
It shows $i_k$ is no longer bounded by $i_{k-1}$ and $i_{k+1}$. The auxiliary matrix $\mathcal{M}$ becomes obsolete because sums of consecutive elements in the superset are no longer needed. We apply the same multithreading technique in \mbox{Section \ref{kopt}} to solving GAP.

For the sake of clarity, consider the following example of assigning 2 tasks to 3 agents:
\begin{equation}
\boldsymbol{c}=\left[\begin{array}{ccc}
21 & 13 & 9\\
6 & 11 & 17
\end{array}\right],\;\;
\boldsymbol{p}=\left[\begin{array}{ccc}
117 & 214 & 167\\
111 & 453 & 20
\end{array}\right],\;\;
\boldsymbol{b}=\left[\begin{array}{ccc}
26 & 25 & 27
\end{array}\right]\;.
\end{equation}
We process the integrated cost and profit matrices by the following steps:
\begin{equation}\label{eq41}
\left[\begin{array}{ccc|c}
21 & 0 & 0 & 117\\
0 & 13 & 0 & 214\\
0 & 0 & 9 & 167\\
\cline{1-4}
6 & 0 & 0 & 111\\
0 & 11 & 0 & 453\\
0 & 0 & 17 & 20
\end{array}\right]%\underset{\text{Order by last column}}{\Rightarrow}
\underset{\parbox{0.08\textwidth}{\scriptsize Order rows by profits in each block}}{\Rightarrow}
\left[\begin{array}{ccc|c}
21 & 0 & 0 & 117\\
0 & 0 & 9 & 167\\
0 & 13 & 0 & 214\\
\cline{1-4}
0 & 0 & 17 & 20\\
6 & 0 & 0 & 111\\
0 & 11 & 0 & 453\\
\end{array}\right]
\underset{\parbox{0.07\textwidth}{\scriptsize Replace with profit ranks}}{\Rightarrow}
\left[\begin{array}{ccc|c}
21 & 0 & 0 & 0\\
0 & 0 & 9 & 1\\
0 & 13 & 0 & 2\\
\cline{1-4}
0 & 0 & 17 & 0\\
6 & 0 & 0 & 1\\
0 & 11 & 0 & 2\\
\end{array}\right]\rightarrow\boldsymbol{x}\;.
\end{equation}
The minimal discrete differential in $\boldsymbol{x}(,0)$, $\boldsymbol{x}(,1)$, $\boldsymbol{x}(,2)$ are -21, -13 and -17 respectively. Instead of taking a different comonotonization multiplier for each column, as Equation \eqref{eq18} suggests, here we take a number larger than the negative of the lowest discrete differential of all columns, e.g. 22, as the universal comonotonization multiplier for all columns. The advantage of such choice is shown below.

Given the comonotonization multiplier 22, Equation \eqref{eq41} undergoes the following transformation:
\begin{equation}\label{eq42}
\boldsymbol{x}\Rightarrow
\left[\begin{array}{ccc|c}
\frac{21 + 0 \times 22}{22} & \frac{0 + 0 \times 22}{22} & \frac{0 + 0 \times 22}{22} & 0\\
\frac{0 + 1\times 22}{22} & \frac{0 + 1\times 22}{22} & \frac{9 + 1\times 22}{22} & 1\\
\frac{0 + 2 \times 22}{22} & \frac{13 + 2\times 22}{22} & \frac{0 + 2\times 22}{22} & 2\\
\cline{1-4}
\frac{0 + 0 \times 22}{22} & \frac{0 + 0 \times 22}{22} & \frac{17 + 0 \times 22}{22} & 0\\
\frac{6 + 1 \times 22}{22} & \frac{0 + 1\times 22}{22} & \frac{0 + 1\times 22}{22} & 1\\
\frac{0 + 2 \times 22}{22} & \frac{11 + 2\times 22}{22} & \frac{0 + 2\times 22}{22} & 2
\end{array}\right]\Rightarrow
\left[\begin{array}{ccc|c}
0.955 & 0 & 0 & 0\\
1 & 1 & 1.409 & 1\\
2 & 2.591 & 2 & 2\\
\cline{1-4}
0 & 0 & 0.773 & 0\\
1.273 & 1 & 1 & 1\\
2 & 2.5 & 2 & 2
\end{array}\right]\rightarrow\boldsymbol{x}\,.
\end{equation}
There are five unique size-2 subset sums in the key column: 4, 3, 2, 1, 0. Five subset sum ranges are thus subject to mining:
\begin{equation}\label{eq43}
\begin{split}
\boldsymbol{S}_L&=
\left[\begin{array}{cccc}
-\infty & -\infty & -\infty & 4\\
-\infty & -\infty & -\infty & 3\\
-\infty & -\infty & -\infty & 2\\
-\infty & -\infty & -\infty & 1\\
-\infty & -\infty & -\infty & 0
\end{array}\right],\;\\
\boldsymbol{S}_U&=
\left[\begin{array}{cccc}
\frac{26 + 4\times 22}{22} & \frac{25 + 4\times 22}{22} & \frac{27 + 4\times 22}{22} & 4\\
\frac{26 + 3\times 22}{22} & \frac{25 + 3\times 22}{22} & \frac{27 + 3\times 22}{22} & 3\\
\frac{26 + 2\times 22}{22} & \frac{25 + 2\times 22}{22} & \frac{27 + 2\times 22}{22} & 2\\
\frac{26 + 1\times 22}{22} & \frac{25 + 1\times 22}{22} & \frac{27 + 1\times 22}{22} & 1\\
\frac{26 + 0\times 22}{22} & \frac{25 + 0\times 22}{22} & \frac{27 + 0\times 22}{22} & 0\\
\end{array}\right]=
\left[\begin{array}{cccc}
5.182 & 5.136 & 5.227 & 4\\
4.182 & 4.136 & 4.227 & 3\\
3.182 & 3.136 & 3.227 & 2\\
2.182 & 2.136 & 2.227 & 1\\
1.182 & 1.136 & 1.227 & 0\\
\end{array}\right]
\end{split}
\end{equation}
where $\boldsymbol{S}_L(s,)$ and $\boldsymbol{S}_U(s,)$, ${s\in[0,\,4]}$ account for one subset sum range.

For row $s$ in every block of $\boldsymbol{x}$, there exists only one element that is fractional and is always greater than $s$. This property is exploited for speedup via compact representation of $\boldsymbol{x}$. More specifically, each row of $\boldsymbol{x}$ is represented by two values: the fraction and its column index. Algorithms for summation, subtraction and comparison are tuned for the compact representation. These algorithms are also considerably easier to implement.

GAP in \pkg{FLSSS} is an exact algorithm. For large-scale suboptimality-sufficient problems, the speed performance may not catch up with fast heuristics such as \citep{salim} and \citep{nauss} that employs a variety of relaxation and approximation techniques to approach suboptima.
%which attack the problem from an integer programming perspective.

\section{Discussion and outlook}
This article introduced algorithms and engineering details for a variety of subset sum problems. The variety includes problem dimensionality, solution quantity, relaxation on target subset sum, constraints on subset size and subset elements. The algorithmic framework applies to the knapsack problem and the generalized assignment problem as exact algorithms. For the package's interest, record-holding heuristics suitable for large-scale and suboptimality-sufficient problems of generalized assignment or multidimensional knapsack may be implemented to the package in the future.

\bibliography{bibfile}
%\bibliography{refs}

\end{document}